\documentclass[a4paper]{jpconf}
\usepackage{citesort}
\usepackage{graphicx}

\begin{document}
\title{A new twist-3 analysis of the single transverse-spin asymmetry
for pion and kaon production at RHIC}

\author{Koichi Kanazawa$^1$ and Yuji Koike$^2$}

\address{$^1$ Graduate School of Science and Technology, Niigata University,
Ikarashi, Niigata 950-2181, Japan}
\address{$^2$ Department of Physics, Niigata University,
Ikarashi, Niigata 950-2181, Japan}

\ead{kanazawa@nt.sc.niigata-u.ac.jp}

\begin{abstract}
We present a new analysis for the single-transverse spin asymmetry (SSA)
for the inclusive pion and kaon productions in the proton-proton
collisions measured at RHIC.  The framework we use is the collinear factorization
in which SSA appears as a twist-3 observable resulting from the multiparton 
correlations in the hadrons.
%
As a relevant multiparton correlation, we
include all the contributions associated with the
quark-gluon correlation functions in the transversely polarized proton,
i.e., those from the soft-gluon-pole (SGP) and the 
soft-fermion-pole (SFP).
%
We have found that the SGP contribution alone is insufficient to describe the observed 
RHIC $A_N$ data and
the inclusion of the SFP contribution plays an important role for the improved description
of the data. 
In particular, $P_T$-dependence of $A_N$ and $A_N$ for $K^-$ have been well
described in our new analysis. 
\end{abstract}

\section{Introduction}

%

The large single transverse-spin asymmetry (SSA) observed 
in the inclusive single-hadron production, $p^\uparrow p\to h X$
($h=\pi,\ K$ etc.), at RHIC\,\cite{Star2004,Star2008,Brahms2008}
plays a crucial role
to reveal the multiparton correlations inside the nucleon.  
%
%
%
In the region of the large transverse momentum of the final hadron, 
$P_{T}\sim Q \gg
\Lambda_{\rm QCD}$, the SSA is described as a twist-3 observable in the
collinear factorization, connecting the SSA to the multiparton correlations directly
\cite{QiuSterman1998,KouvarisEtal2006,EguchiKoikeTanaka2006,EguchiKoikeTanaka2007,KoikeTomita2009,KangYuanZhou2010}.
In principle, multiparton correlations causing the SSA in 
$p^\uparrow p\to h X$ exist both in the initial-state proton and the final-state
hadron.  
%
Since the analytic formula for the latter effect is not available,  
we shall focus in this work on the effect of the quark-gluon correlations in the
transversely polarized proton and see how the existing data are described 
by those effects. \footnote{The twist-3 fragmentation function for the final hadron is chiral-odd
and appears in pair with the transversity distribution.  We assume this chiral-odd
effect may be small in this initial study.}
In the transversely polarized nucleon, 
there are
two independent twist-3 quark-gluon correlation functions, 
$G_F(x_1,x_2)$ and
$\widetilde{G}_F(x_1,x_2)$, which are defined from 
the light-cone correlation function in the nucleon 
$\sim \langle \bar{\psi} F^{\alpha +} \psi
\rangle$, with $\psi$ being the quark field and $F^{\alpha +}$
the gluon's field strength. 
(For the explicit definition of the correlation functions and their property, see
\cite{EguchiKoikeTanaka2006,EguchiKoikeTanaka2007}.) 
The variables $x_{1,2}$ and $x_2-x_1$ denote, respectively, the 
longitudinal momentum fractions of the quarks and the gluon coming out of the nucleon.  
In the twist-3 mechanism, the complex phase necessary for SSA is supplied
as a pole contribution from an internal propagator in the hard part.
These poles
are classified into the soft-gluon-pole (SGP) and soft-fermion-pole (SFP),
corresponding, respectively, to $x_i=0$ ($i=1$ or 2) and $x_2-x_1=0$.  
Accordingly, the single-spin-dependent cross section 
can be represented as
\begin{eqnarray}
 \Delta\sigma^{\rm tw3}_{p^{\uparrow}p\to hX} &\sim& \left( G_F(x,x)-x\frac{dG_F(x,x)}{dx} \right) \otimes f(x')
 \otimes D(z) \otimes \hat{\sigma}_{\rm SGP} \nonumber\\
 &+& \left( G_F(0,x)+\widetilde{G}_F(0,x) \right) \otimes f(x') \otimes D(z) \otimes
 \hat{\sigma}_{\rm SFP} \label{formula},
\end{eqnarray}
where the first and the second term represents, respectively, the SGP and the SFP contributions, and
$f(x')$ and $D(z)$ are the unpolarized parton distribution and the fragmentation
functions.
Kouvaris {\it et al.} \cite{KouvarisEtal2006} derived $\hat{\sigma}_{\rm SGP}$
in (\ref{formula}) and performed a numerical analysis
of the existing $A_N$ data. \footnote{For the origin of the combination $G_F(x,x)-x\frac{dG_F(x,x)}{dx}$
for the SGP contribution, see \cite{KoikeTanaka2007}.}     
They obtained a reasonable description of the RHIC data except for the positive
asymmetry for the $K^-$ production and the $P_T$-dependence. 
One of the present authors derived the formula for $\hat{\sigma}_{\rm SFP}$\,\cite{KoikeTomita2009} and  
found $\hat{\sigma}_{\rm SFP} \gg \hat{\sigma}_{\rm SGP}$ in the relevant channels,
showing a potential importance of the SFP contribution. 
%
%

%
In our recent paper\,\cite{KanazawaKoike2010}, we have analyzed the
RHIC $A_N$ data based on the
formula (\ref{formula}) including both SGP and SFP contributions.
We found that the inclusion of the SFP contribution gives a better description of the
data and that 
the SFP contribution can not be replaced
by the readjustment of the SGP contribution.
This is the first numerical analysis showing the importance of the 
SFP effect.  This talk provides a short summary of the analysis and 
some new plots not shown in \cite{KanazawaKoike2010}.    

\section{Phenomenological analysis of RHIC $A_N$ data}

\subsection{Fitting}

For the fitting, 
we have assumed the following functional form for the SGP and
SFP functions,
\begin{eqnarray}
&& G_{F,a}(x,x) = N_a^G x^{\alpha^G_a} (1-x)^{\beta_a^G} f_a(x) \nonumber\\
&& G_{F,a}(0,x) + \widetilde{G}_{F,a}(0,x) = N_a^F x^{\alpha_a^F}
 (1-x)^{\beta_a^F} f_a(x) , \nonumber
\end{eqnarray}
where the subscript $a$ denote the light-quark flavor, $a=u,d,s,\bar{u},\bar{d},\bar{s}$, and 
$N_a^{G,F},\alpha_a^{G,F}$ and $\beta_a^{G,F}$ are the parameters to be
determined by the fitting.  For simplicity, the scale dependence of each
twist-3 functions
are assumed to be the same as the unpolarized quark distribution function
$f_a(x)$ in the
right-hand-side. In our fitting, we have included the RHIC data by the
STAR\,\cite{Star2004,Star2008} ($\sqrt{S}=200$ GeV) and 
the BRAHMS\,\cite{Brahms2008} ($\sqrt{S}=62.4$ GeV) collaborations.
We did not use the FNAL data ($\sqrt{S}=20$ GeV), since the NLO QCD 
in the collinear factorization fails to reproduce the unpolarized cross section
at the fixed target energy.

Under this assumption, we have performed 
three types of fitting in order to see the importance of the SFP effect.   
FIT 1 includes both SGP and SFP contributions.  FIT 2 omit the SFP contribution
from FIT 1 with the same degrees of freedom as FIT 1 for the SGP contribution. 
FIT 3 includes only the SGP contribution but with more degrees of freedom for the parameters.
For the fits, we used the GRV98 \cite{GlueckReyaVogt1998} parton distribution and the DSS fragmentation
functions for $\pi$ and $K$ \cite{FlorianSassotStratmann2007PIK}.  
\begin{figure}[h]

\begin{center}

\scalebox{0.65}{\includegraphics{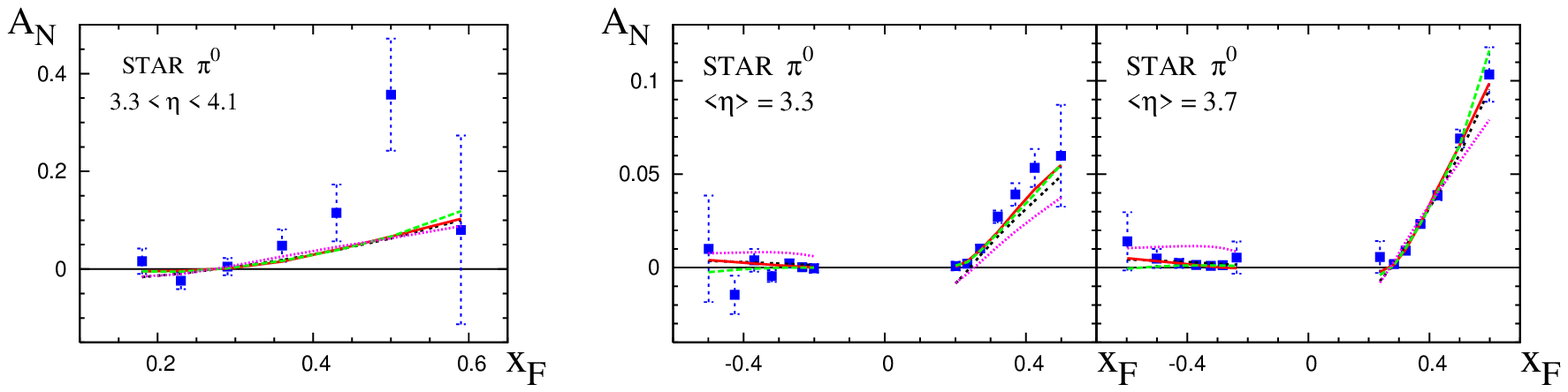}}

\scalebox{0.65}{\includegraphics{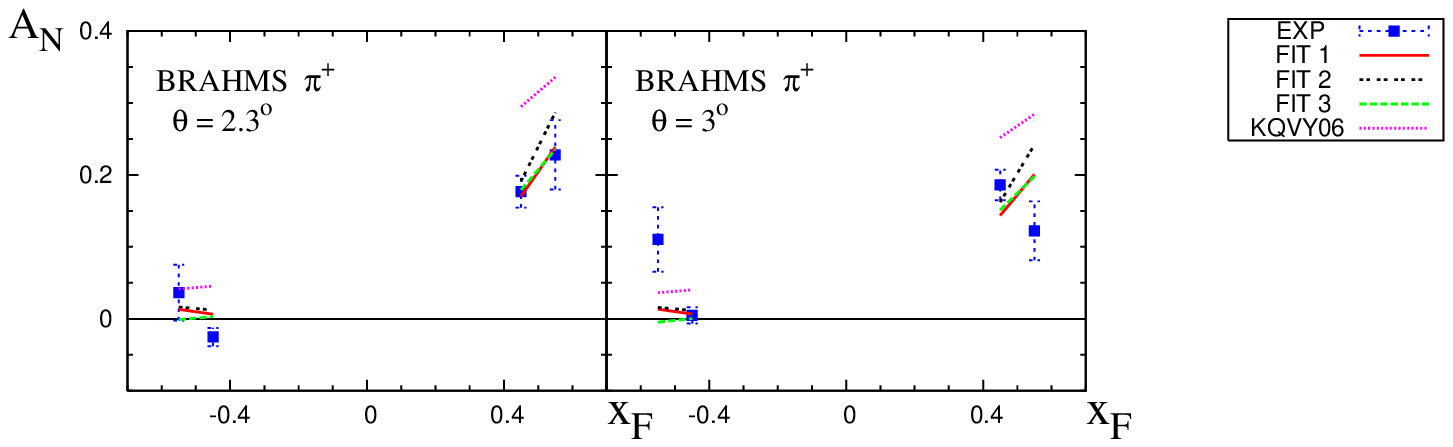}}

\end{center}

\begin{minipage}{0.5\hsize}

\scalebox{0.65}{\includegraphics{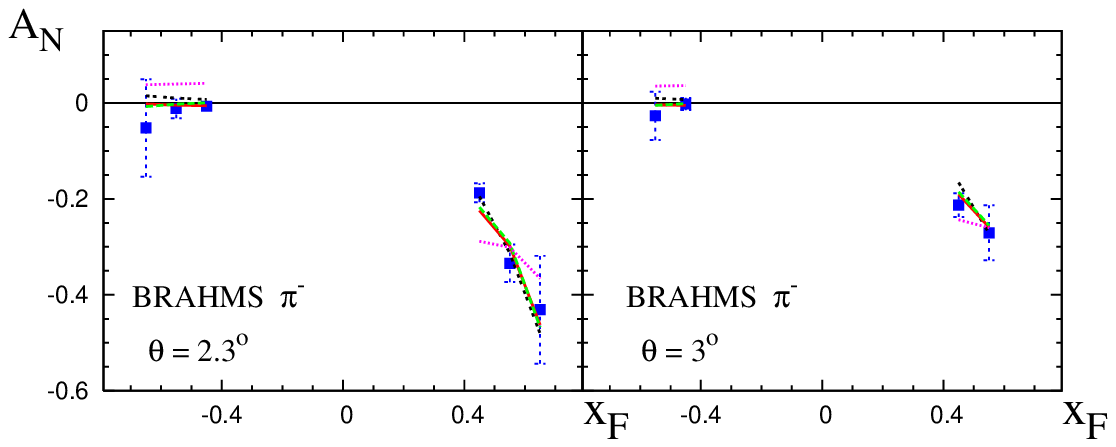}}

\end{minipage}
\begin{minipage}{0.5\hsize}

\scalebox{0.65}{\includegraphics{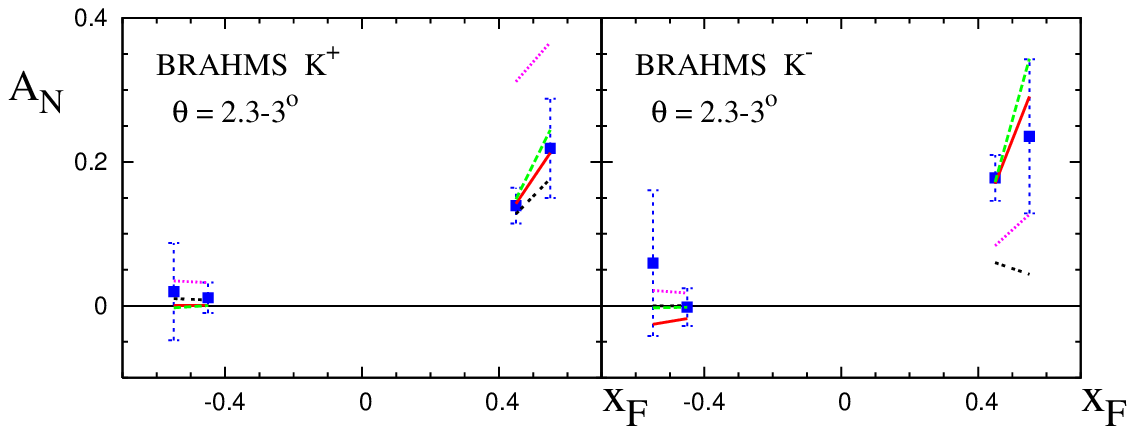}}

\end{minipage}

\caption{Results of the analysis in comparison with the RHIC data \cite{Star2004,Star2008,Brahms2008}.\label{result}}

\end{figure}
\begin{figure}[h]
 \begin{center}
  \begin{minipage}{0.3\hsize}
   \scalebox{0.7}{\includegraphics{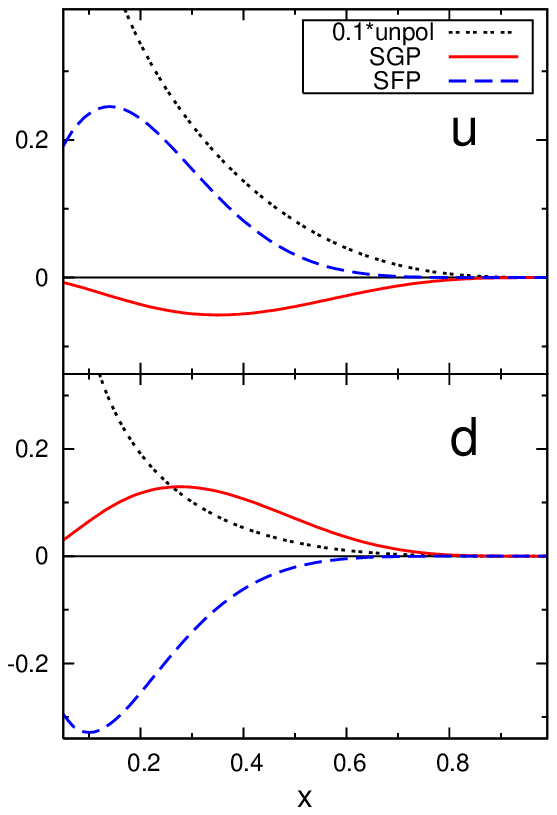}}
  \end{minipage}
  \begin{minipage}{0.3\hsize}
   \scalebox{0.7}{\includegraphics{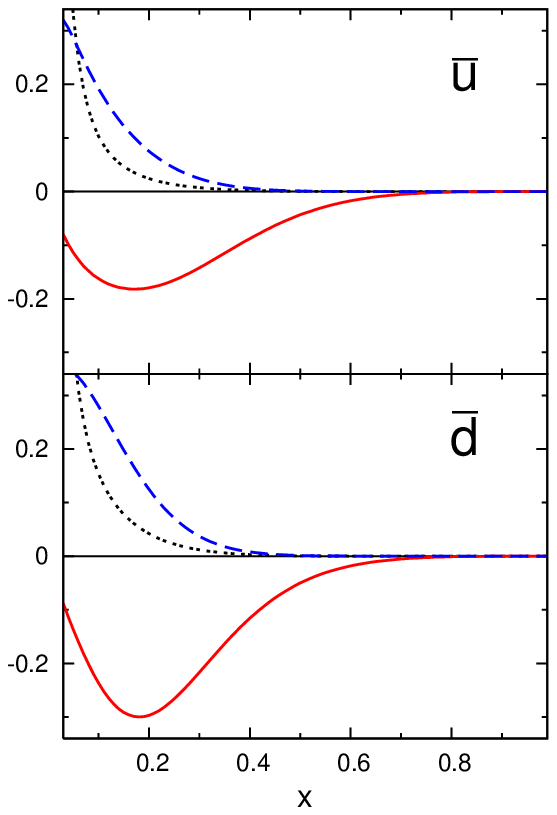}}
  \end{minipage}
  \begin{minipage}{0.3\hsize}
   \scalebox{0.7}{\includegraphics{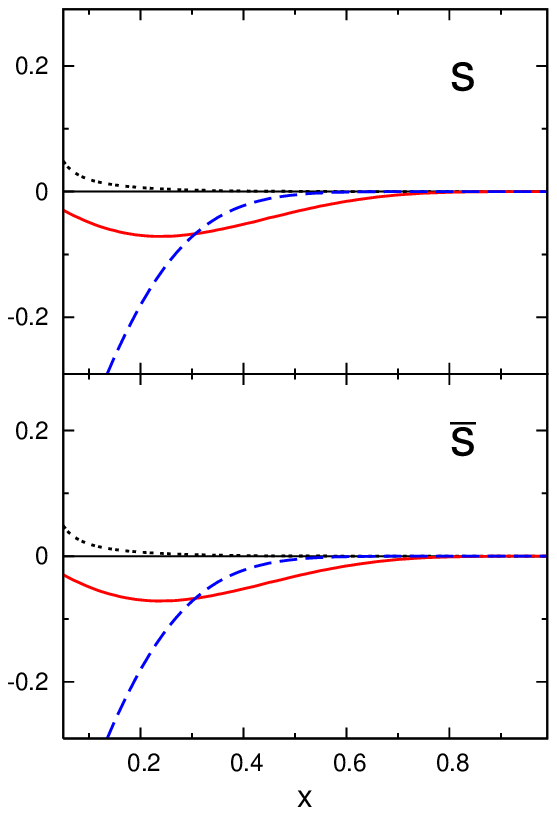}}
  \end{minipage}
  \caption{The obtained SGP and SFP functions in FIT 1 in comparison with the
  unpolarized parton distribution function scaled by $1/10$. \label{sgp}}
 \end{center}
\end{figure}
The obtained results are shown in Fig. \ref{result} together with the RHIC data.  For comparison 
we also showed the result by Kouvaris {\it et al.} (Fit II of \cite{KouvarisEtal2006}).
One sees that FIT 1 and FIT 3 give good agreement with the data, while FIT 2
fails to reproduce $A_N$ for $K^-$ production.
This feature is reflected in the values of $\chi^2/d.o.f$, which are
$1.21$,  $2.46$ and $1.32$ for FIT 1, 2 and 3, respectively.  
%
Although FIT 3 reproduce the data, some of the obtained SGP functions in FIT 3 
show an unphysical behavior, which causes large $A_N$ for $\pi$ in the small $x_F$ region
at the FNAL energy.  
From these facts, we regard our FIT 1 as the best fit for the RHIC $A_N$ data.   
%
%
%

Fig. 2 shows the SGP and SFP functions for each quark flavor in FIT 1.  
One sees from the figure that the SGP and the SFP functions have similar magnitude, but
the SGP functions spreads more in the larger $x$ region, which plays an important role 
for the rising $A_N$ at large $x_F$.  Among the SFP functions, 
those for the $u$ and $d$ quarks are larger and have longer range than those for other flavors.

\subsection{$P_{T}$-dependence of the asymmetry}

\begin{figure}[h]
 \begin{minipage}{0.5\hsize}
  \begin{flushleft}
   \scalebox{0.6}{\includegraphics{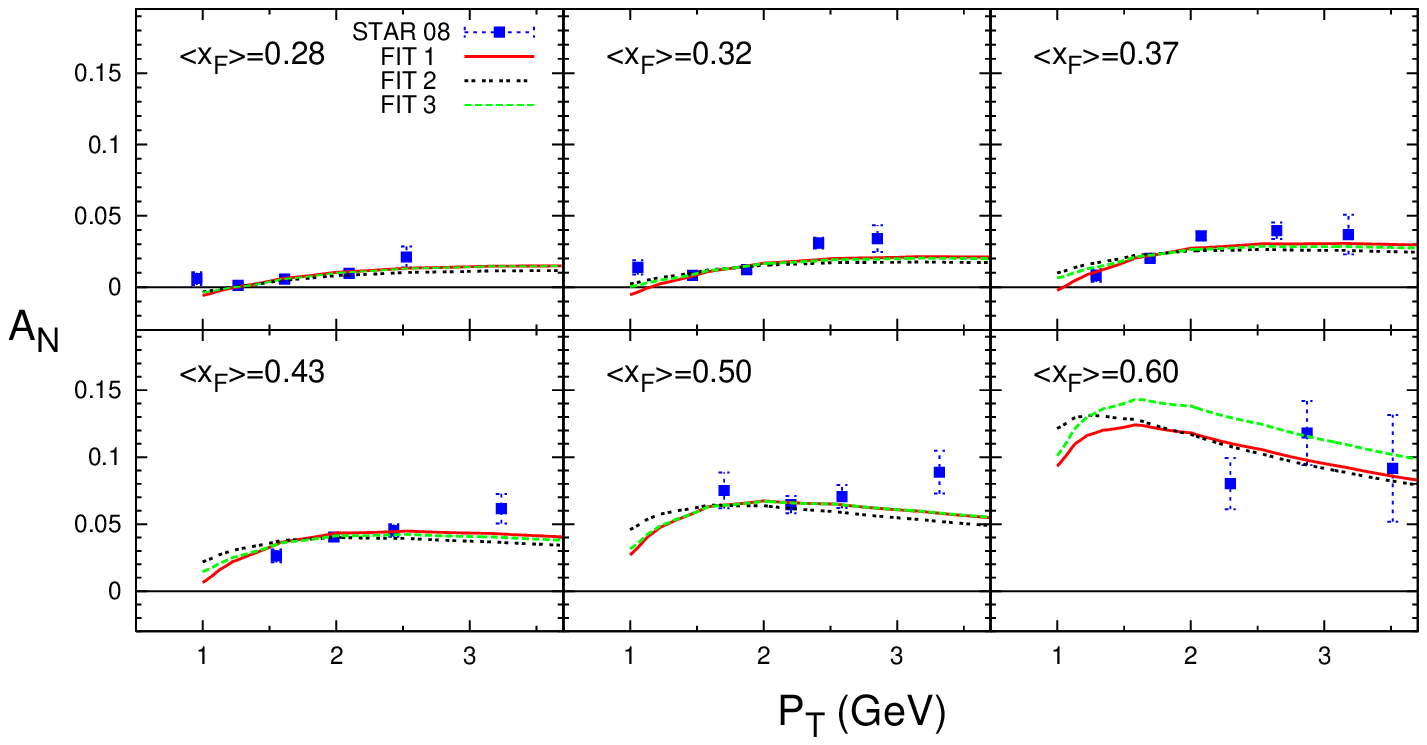}}
  \end{flushleft}
 \end{minipage}
 \begin{minipage}{0.5\hsize}
  \begin{center}
   \scalebox{0.83}{\includegraphics{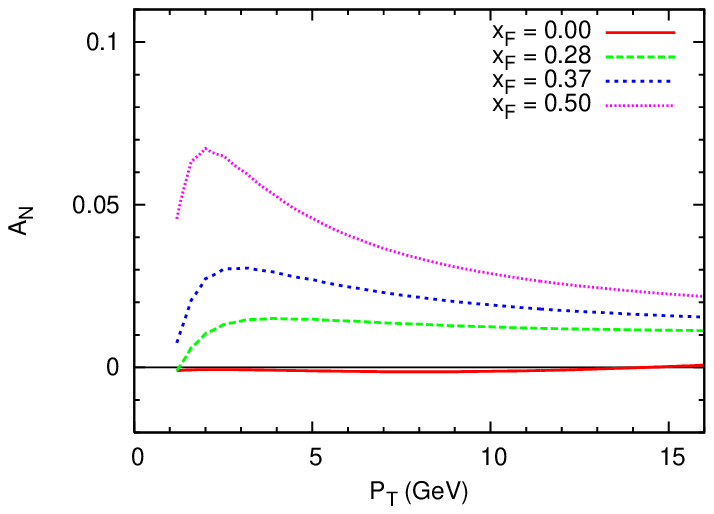}}
  \end{center}
 \end{minipage}
\caption{$P_{T}$-dependence of $A_N^{\pi^0}$ compared with
 the STAR data \cite{Star2008} (left), and the $P_T$-dependence of
 $A_N^{\pi^0}$ in the large $P_T$ region for FIT 1 (right).\label{pt}}
\end{figure}

Another interesting feature of $A_N$ is its dependence on the transverse momentum ($P_T$)
of the final hadron.  
Fig. \ref{pt} shows the calculated $P_T$-dependence of $A_N$ for $\pi^0$
together with the STAR data (left figure) and that in the wider range of $P_T$
for FIT 1 (right figure).   All three fits reproduce the $P_T$-dependence
of the data.  This is quite natural since our fit used the 
corresponding $P_T$ values for each data point shown in Fig. 1, and thus
the $P_T$-dependence shown in Fig. \ref{pt} is partly taken into account.  
At larger $P_T$, $A_N$ starts decreasing but not as fast as $\sim M_N/P_T$.
This is because two effects of $O\left( M_NP_T/(-T)\right)$ and $O\left( M_NP_T/(-U)\right)$ 
coexist in $A_N$ reflecting its twist-3 nature, which leads to rather flat $P_T$-behavior
at moderate values of $x_F$.  
At $x_F=0$, $A_N$ stays zero in the whole $P_T$ region which is consistent with the
PHENIX data\,\cite{Phenix2005}.

\subsection{Role of the SFP contribution and the flavor decomposition}

By separating the calculated $A_N$ into the SGP and the SFP contributions,
we found that a major contribution comes from the former effect.  However, 
the SFP contribution also brings a large contribution at $x_F\leq 0.3$ which 
cancels the SGP contribution, making 
$A_N$ small in this region.  For $K^-$ and $\pi^-$, we found a significant contribution
from the SFP effect.  For the detail, see Fig. 4 in \cite{KanazawaKoike2010}.

\begin{figure}[h]
\begin{center}

{\large $\pi^+$}

\scalebox{0.7}{\includegraphics{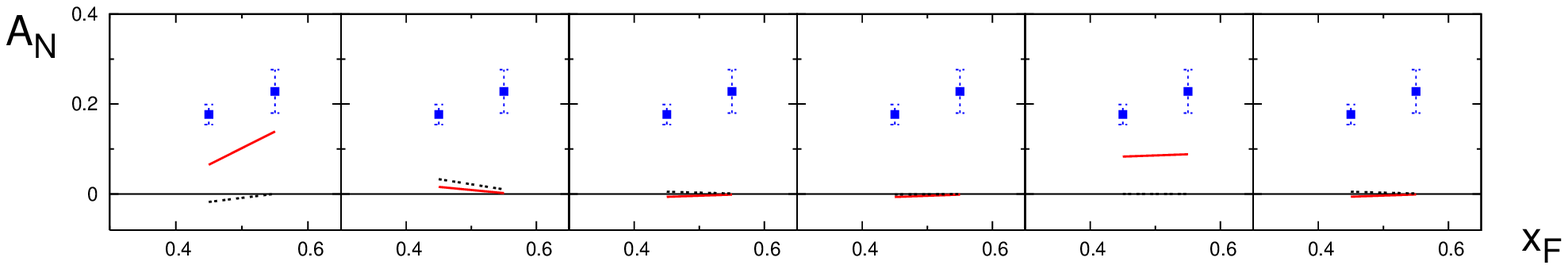}}

\vspace{4pt}

{\large $\pi^-$}

\scalebox{0.7}{\includegraphics{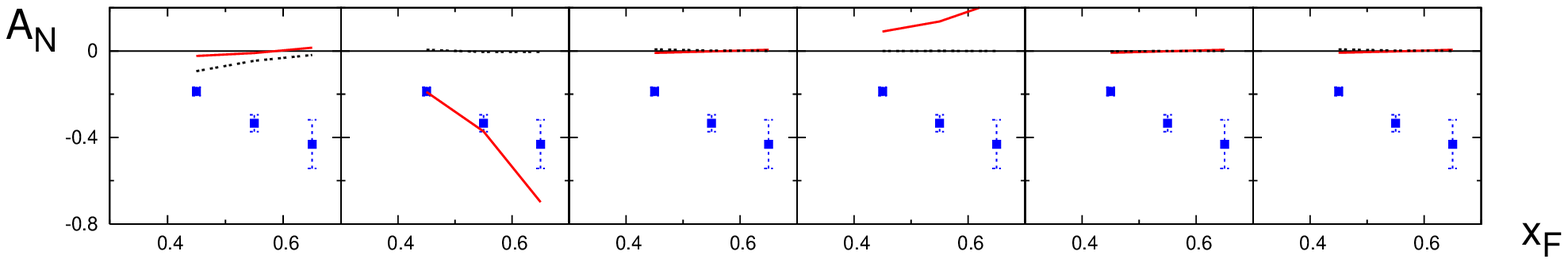}}

\vspace{4pt}

{\large $K^+$}

\scalebox{0.7}{\includegraphics{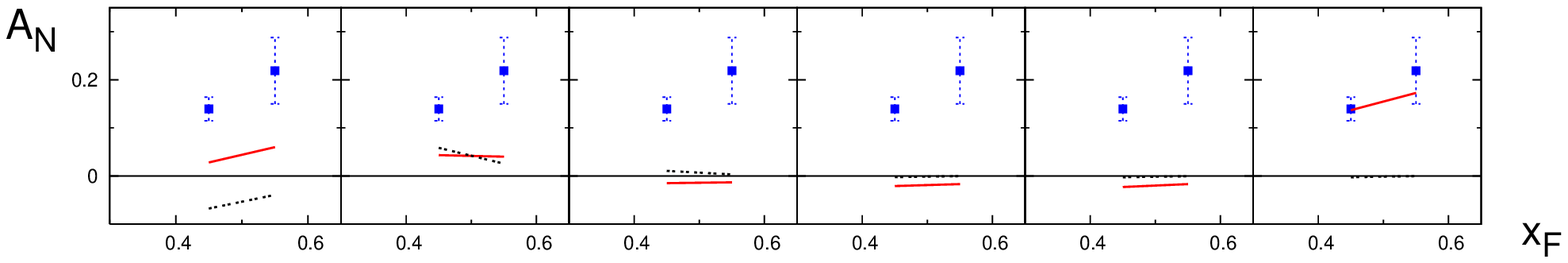}}

\vspace{4pt}

{\large $K^-$}

\scalebox{0.7}{\includegraphics{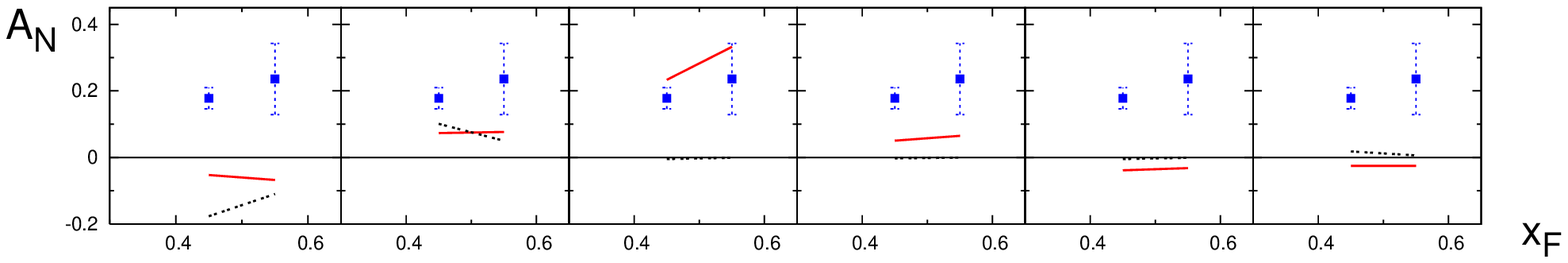}}

\vspace{4pt}

\begin{tabular}{c}
{\large Flavor : \hspace{10pt} $u$ \hspace{42pt} $d$ \hspace{42pt} $s$ \hspace{42pt} $\bar{u}$
 \hspace{42pt} $\bar{d}$ \hspace{42pt} $\bar{s}$ \hspace{55pt}} \\
\hline
\end{tabular}

\caption{Flavor decomposition of $A_N$ for the charged  meson production
 in FIT 1. The solid and dashed lines show the
 contributions from the SGP and SFP components, respectively. \label{fd}}

\end{center}
\end{figure}

To see more detail of FIT 1, we have shown in Fig. \ref{fd}
the decomposition of 
$A_N$ at $x_F\ge 0.4$ for $\pi^\pm$ and $K^\pm$ into each quark-flavor
of the quark-gluon correlation functions.  
As is seen from the figure, the dominant contribution comes from the SGP effect
for the valence flavor.  However, for $K^\pm$, the SFP effect from
$u$- and $d$-quark flavors is also significant.  This is because of the corresponding 
large SFP functions (see Fig. 2) and the 
large partonic hard cross section in the gluon-fragmentation channel
combined with the large gluon component for $K$ in the DSS fragmentation function. 
For the pion, this SFP effect in the gluon-fragmentation channel 
is canceled by the quark contributions, making the net SFP effect small.

\section{Conclusion and outlook}

In this work we have presented a new analysis of the RHIC $A_N$ data 
for $\pi$ and $K$
based on the
twist-3 mechanism, including the complete contributions from the quark-gluon
correlation functions in the transversely polarized nucleon.
We have found that the SGP contribution alone is insufficient 
and the inclusion of the SFP contribution is crucial to reproduce
all the reported RHIC data. 
%
%
The SGP and the SFP functions determined in our analysis 
can be used to predict the SSA in other processes such as the
direct-photon production and 
Drell-Yan process at RHIC, and semi-inclusive deep inelastic scattering at EIC energy, and should 
be tested by confrontation with experimental data.  
%

In our present study, other twist-3 effects, i.e., the three-gluon correlations in the
transversely polarized proton 
and the
contribution from the twist-3 fragmentation function, have not been taken into
account.  
For a complete clarification of the origin of SSA, these effects should
also be investigated
by looking at wider variety
of experimental data in the future study.
%

\vspace{11pt}

\section*{References}

\providecommand{\newblock}{}


\begin{thebibliography}{10}
\expandafter\ifx\csname url\endcsname\relax
  \def\url#1{{\tt #1}}\fi
\expandafter\ifx\csname urlprefix\endcsname\relax\def\urlprefix{URL }\fi
\providecommand{\eprint}[2][]{\url{#2}}

\bibitem{Star2004}
Adams J {\em et~al.\/} (STAR Collaboration) 2004 {\em Phys. Rev. Lett.\/} {\bf
  92} 171801

\bibitem{Star2008}
Abelev B~I {\em et~al.\/} (STAR Collaboration) 2008 {\em Phys. Rev. Lett.\/}
  {\bf 101} 222001

\bibitem{Brahms2008}
Arsene I {\em et~al.\/} (BRAHMS Collaboration) 2008 {\em Phys. Rev. Lett.\/}
  {\bf 101} 042001

\bibitem{QiuSterman1998}
Qiu J and Sterman G 1998 {\em Phys. Rev.\/} D {\bf 59} 014004

\bibitem{KouvarisEtal2006}
Kouvaris C, Qiu J~W, Vogelsang W and Yuan F 2006 {\em Phys. Rev.\/} D {\bf 74}
  114013

\bibitem{EguchiKoikeTanaka2006}
Eguchi H, Koike Y and Tanaka K 2006 {\em Nuclear Physics\/} B {\bf 752} 1

\bibitem{EguchiKoikeTanaka2007}
Eguchi H, Koike Y and Tanaka K 2007 {\em Nuclear Physics\/} B {\bf 763} 198

\bibitem{KoikeTomita2009}
Koike Y and Tomita T 2009 {\em Physics Letters\/} B {\bf 675} 181

\bibitem{KangYuanZhou2010}
Kang Z~B, Yuan F and Zhou J 2010 {\em Physics Letters\/} B {\bf 691} 243

\bibitem{KoikeTanaka2007}
Koike Y and Tanaka K 2007 {\em Phys. Rev.\/} D {\bf 76} 011502

\bibitem{KanazawaKoike2010}
Kanazawa K and Koike Y 2010 {\em Phys. Rev.\/} D {\bf 82} 034009

\bibitem{GlueckReyaVogt1998}
Gl\"{u}ck M, Reya E and Vogt A 1998 {\em Eur. Phys. J.\/} C {\bf 5} 461

\bibitem{FlorianSassotStratmann2007PIK}
de~Florian D, Sassot R and Stratmann M 2007 {\em Phys. Rev.\/} D {\bf 75}
  114010

\bibitem{Phenix2005}
Adler S~S {\em et~al.\/} (PHENIX Collaboration) 2005 {\em Phys. Rev. Lett.\/}
  {\bf 95} 202001

\end{thebibliography}

\end{document}